\documentclass[preprint]{aastex}
\shorttitle{Source Imagery of a T-IVm Solar Radio Burst}
\shortauthors{Vasanth et al.}
\usepackage{graphicx}
\usepackage{xcolor}
\begin{document}

\title{Source Imaging of a Moving Type-IV Solar Radio Burst and its Role in Tracking Coronal Mass Ejection From the Inner to the Outer Corona}

\author{V. Vasanth\altaffilmark{1}, Yao Chen\altaffilmark{1}, Maoshui Lv\altaffilmark{1}, Hao Ning\altaffilmark{1},
Chuangyang Li\altaffilmark{1}, Shiwei Feng\altaffilmark{1}, Zhao Wu\altaffilmark{1}, and Guohui Du\altaffilmark{1}}

\altaffiltext{1}{Shandong Provincial Key Laboratory of Optical
Astronomy and Solar-Terrestrial Environment, and Institute of
Space Sciences, Shandong University, Weihai 264209, China;
yaochen@sdu.edu.cn, vasanth@sdu.edu.cn}

\begin{abstract}
Source imaging of solar radio bursts can be used to track energetic electrons and associated magnetic structures. Here we present a combined analysis of data at different wavelengths for an eruption associated with a moving type-IV (t$\--$IVm) radio burst. In the inner corona, the sources are correlated with a hot and twisted eruptive EUV structure, while in the outer corona the sources are associated with the top front of the bright core of a white light coronal mass ejection (CME). This reveals the potential of using t$\--$IVm imaging data to continuously track the CME by lighting up the specific component containing radio-emitting electrons. It is found that the t$\--$IVm burst presents a clear spatial dispersion with observing frequencies. The burst manifests broken power-law like spectra in brightness temperature, which is as high as $10^7 $\--$ 10^9$~K while the polarization level is in-general weak. In addition, the t-IVm burst starts during the declining phase of the flare with a duration as long as 2.5 hours. From the differential emission measure analysis of AIA data, the density of the T-IVm source is likely at the level of 10$^8$~cm$^{-3}$ at the start of the burst, and the temperature may reach up to several MK. These observations do not favor gyro-synchrotron to be the radiation mechanism, yet in line with a coherent plasma emission excited by energetic electrons trapped within the source. Further studies are demanded to elucidate the emission mechanism and explore the full diagnostic potential of t-IVm bursts.
\end{abstract}

\keywords{Sun: corona --- Sun: activity --- Sun: coronal mass
ejections (CMEs) --- Sun: radio radiation}

\section{INTRODUCTION}
Coronal Mass Ejections (CMEs) release a vast amount of fast moving magnetized plasmas into the interplanetary space and may cause catastrophic space weather effects when propagating towards the Earth. To understand the underlying physics of CMEs and develop forecasting methods, it is crucial to observe the complete evolutionary process of an eruption, including its initiation (i.e., the triggering and early acceleration process) in source regions from the solar disk to the inner corona, further acceleration to the outer corona, and its propagation into the interplanetary space.

From the solar disk to the inner corona, the structures can be observed at various Extreme Ultraviolet (EUV) wavelengths, covering a broad range of temperatures of emitting plasmas; while from the inner to the outer corona, the eruption is usually observed with coronagraphs in white light (WL). Present EUV imaging instruments include the Atmospheric Imaging Assembly (AIA: Lemen et al. 2012) onboard the \emph{Solar Dynamics Observatory} (\emph{SDO}: Pesnell, Thompson and Chamberlin et al. 2012) and the Sun Watcher using Active Pixel System Detector and Image Processing (SWAP: Seaton et al. 2013; Halain et al. 2013) onboard the second \emph{Project for Onboard Autonomy} (\emph{PROBA2}: Santandrea et al. 2013) and the EUV Imagers (EUVI) onboard the twin \emph{Solar Terrestrial Relations Observatory} (\emph{STEREO}; Kaiser et al. 2008). The nominal field of view (FOV) of AIA extends up to 1.2 R$_\odot$ and that of SWAP extends to $\sim$ 1.6 R$_\odot$. Mainly due to rapidly-declining plasma density with height, at the outer portion of their FOV, the data signal$\--$noise ratio decreases rapidly. On the other hand, the FOV of present space-based WL coronagraphs along the Sun$\--$Earth line, such as the Large Angle Spectrometric Coronagraph (LASCO: Brueckner et al. 1995) C2 onboard the \emph{Solar and Heliospheric Observatory} (\emph{SOHO}: Domingo, Fleck and Poland 1995), starts from 2.2 R$_\odot$ (The C1 coronagraph with a lower FOV of 1.1 $\--$ 3 R$_\odot$ was lost during the early stage of the \emph{SOHO} mission). Note that the twin STEREO spacecraft orbits around the Sun, one ahead and the other behind the Earth, separated from each other by $\sim$ 44 degrees per year. Thus, along the Sun-Earth line a gap exists in regions of $\sim$ 1.2 $\--$ 2 R$_\odot$ between space-based EUV and WL CME data.

Within this region essential evolution of CMEs, such as rapid acceleration, super-radial expansion, strong interaction with nearby structures (e.g., helmet streamers, see, e.g., Chen et al. 2010, 2011; Feng et al. 2011), and morphological transformation, do take place. This contributes to difficulties in resolving several CME$\--$related issues. For instance, CMEs sometimes manifest a three$\--$component structure, consisting of a bright front (double fronts if a shock wave exists), a dark cavity, and a bright core. It remains elusive regarding the exact nature of these components, especially the underlying magnetic structure of the cavity and the core. It has been suggested that the cavity corresponds to an over$\--$expanding (thus less dense) magnetic flux rope structure while the core corresponds to dense filament material carried outward by the flux rope. However, it has been found that a bright CME core can appear without obvious ejection of filament materials (Howard et al. 2017; Song et al. 2017). Thus, it is critical to develop methods that can bridge the mentioned gap of EUV and WL data, to identify the counterparts of each component and track their evolution.

In addition to EUV and WL observations, CMEs can also be observed at metric to decametric radio wavelengths. Among the five types of solar radio bursts (Wild \& McCready 1950; Melrose 1980), two types are directly associated with outward-propagating coronal structures. They are Type-IIs (T-IIs), slowly-drifting narrow bands sometimes present fundamental and harmonic branches as well as splitting bands on the radio dynamic spectrum (Payne-Scott, Yabsley \& Bolton 1947; Wild et al. 1963; see Kong et al. 2012; Feng et al. 2013; Chen et al. 2014; Vasanth et al. 2014; Du et al. 2015 for latest studies), and moving T-IV bursts (T-IVms), broadband continuum with sources moving outward from the corona and frequencies declining with time gradually (Boischot, 1958; Riddle 1970; Sheridan 1970; See Tun \& Vourlidas 2013; Bain et al. 2014; Vasanth et al. 2016; Carley et al. 2017 for latest studies). T-IIs are found to be associated with the shock driven by the eruption (Uchida 1960), and T-IVm bursts manifest different source geometries such as plasmoid, advancing front, or loop-like structure (Smerd and Dulk 1971; Wild \& Smerd 1972; Valhos et al. 1982; Stewart 1985), indicating they are associated with various parts of the ejecta.

Earlier studies of T-IVm sources are mostly based on radioheliographs such as Culgoora at a few discrete frequencies (40, 80, and 160 MHz, see, e.g., Wild 1970; Smerd \& Dulk 1971; Kai 1978). T-IVms are found to be closely associated with corona transients, including eruptive prominence and flare surges (e.g., Wild, Sheridan, \& Kai 1968; Riddle 1970; Sheridan 1970; Riddle et al. 1974; Stewart et al. 1982), even before CMEs were discovered by space-based coronagraph on board the Orbiting Solar Observatory (Hansen et al. 1971; Tousey 1973). Yet, it remains elusive of the exact correspondence between radio$\--$emitting structure and CME component observed at EUV and WL, partially due to scarcity of simultaneous imaging data at different wavelengths.

The emission mechanism of T-IVms is a controversial issue. Both incoherent gyrosynchrotron emission and coherent plasma emission have been suggested. According to earlier studies using one-dimensional interferometric and/or two-dimensional radioheliograph data, T-IV sources can be traced as far as $\sim$5 R$_\odot$ at 80 MHz (e.g., Riddle et al. 1970; Riddle et al. 1974). Such source heights were considered to be much higher than the corresponding plasma level (Smerd \& Dulk, 1971). In addition, no dispersion of T-IV sources with observing frequencies was found in several studies using interferometric data (Wild et al. 1959; Boischot \& Clavelier 1968; Dulk 1973). These observations were considered to be against the plasma emission (Ginzburg \& Zheleznyakov 1958) to be the mechanism of t-IVms, and in line with gyrosynchrotron radiation from mildly relativistic electrons (Boishot 1957; Dulk et al. 1978).

Nevertheless, other studies using the Culgoora radioheliograph data showed that T-IVm sources do exhibit dispersion with observing frequencies (McLean 1973; Nelson 1977; Duncan 1980a; Duncan 1980b), the brightness temperatures ($T_B$) sometimes exceeding 10$^9$ to 10$^{10}$~K, too high to be explained with gyrosynchrotron (Stewart et al. 1978; Duncan 1980a, 1980b; Duncan et al. 1980), and T-IVm sources could be associated with plasmoid structure that is much denser than background plasmas (e.g., Stewart et al. 1982) though the method to infer coronal density suffered from large uncertainty. These observations support that T-IVs are attributed to coherent plasma emission.

In latest studies using the \emph{Nan\c{c}ay Radioheliograh} (NRH: Kerdraon and Delouis 1997) data, Bastain et al. (2001) reported CME-associated radio sources of a moving t-IV burst at metric wavelengths. They found that the radio sources appear as a large-scale expanding loop structure. At a single frequency, the source manifests a spatial extent of more than $\sim$ 4 R$_\odot$, and the sources at different frequencies are co-spatial with each other, i.e., without spatial dispersion with observing frequencies. With state-of-the-art imaging instruments at various passbands, one t-IVm event dated on 14 August 2010 was analyzed by Tun {\&} Vourlidas (2010) and Bain et al. (2014). They reported that the sources present no spatial dispersion at different frequencies and the intensities are in-general weak, and they are associated with the CME core that is likely transformed from an eruptive filament. The emission mechanism of these events was still considered to be gyrosynchrotron.

With another event, Vasanth et al. (2016) showed that t$\--$IVm sources are associated with eruptive structure observed by AIA at its hot channels (131 and 94~\,\AA{}). They found that the sources move outward together with the eruptive structure that is likely a twisted flux rope (see also, Cheng et al. 2011; Zhang et al. 2012). At different frequencies (from 150 $\--$ 445 MHz), the sources first present no considerable spatial dispersion, then spread gradually along the eruptive structure with significant spatial dispersion. Note that earlier studies on t-IV sources mainly using the Culgoora data at lower frequencies (40, 80, and 160 MHz). Vasanth et al. (2016) proposed that the T-IV burst is attributed to coherent emission driven by energetic electrons trapped within the source structure where a positive-gradient of velocity distribution function along the perpendicular direction develops. This is in agreement with several earlier studies (e.g., Winglee {\&} Dulk 1986; Benz 2002). To further discern the emission mechanism, it is necessary to examine more events and infer source properties such as plasma density and temperature, and strength and geometry of magnetic field.

Here we investigate a type$\--$IVm burst. It is unique that there is no interference from other types of solar radio bursts. The burst and the associated solar eruption are well observed by a tandem of instruments. In particular, the multi-wavelength data from AIA/SDO allow us to infer plasma density and temperature within the source structure. These observations, together with the limb perspective, make the event a nice candidate for studies on physical mechanism underlying the t$\--$IVm radio burst and its role in tracking the source continuously from the inner to the outer corona.

\section{Observational Data and Event Overview}

The dynamic spectrum of the t$\--$IVm radio burst was recorded on 2014 June 15 by the ORFEES radio spectrograph from 300$\--$144 MHz and by the Nancy Decametric Array (NDA; see Lecacheux 2000) from 80$\--$10 MHz. The composite of the spectral data is presented in Figure 1. We checked many other radio spectrographs and found no clear signatures of this radio burst. This is likely due to the relatively low intensity of the burst as well as the in-general low sensitivity of those instruments. A data gap exists between 144 - 80 MHz due to the strong radio interference. The burst starts from 12:36 UT at 327 MHz, with an overall drifting trend toward lower frequencies. It lasts for at least 2.5 hours, and reaches the lowest frequency of NDA ($\sim$ 10 MHz, the nominal ionospheric cut-off frequency). The burst is wide-band, without the usual interference of other types of radio bursts, such as the type-II, III, among others. This provides a rare opportunity for the investigation on type-IV radio bursts.

The sources of the burst were imaged by NRH at several frequencies (327, 298, 270, 173, and 150 MHz). The spatial resolution of NRH images depends on the imaging frequency and observational time, being 2\arcmin\ \--3\arcmin\ at 327 MHz, decreasing to 5\arcmin\ \-- 8\arcmin\ at 150 MHz. The highest temporal resolution of NRH is 0.25s, here the data with a 10s resolution are used for higher signal to noise ratio.

The t$\--$IVm burst is associated with a solar eruption from the NOAA active region (AR12085) at the southwestern limb. The eruption is associated with an M1$\--$class flare at around S22W89, which starts at 11:10 UT and peaks at 11:40 UT according to the GOES SXR data (see Figure 1). It involves a CME that is observed by several instruments along the Sun-Earth line, including the AIA/\emph{SDO}, SWAP/\emph{PROBA2}, and LASCO/\emph{SOHO}. The CME was also observed by the EUVI and COR 1/2 instruments onboard the twin \emph{STEREO} A/B spacecraft from different perspectives. For this event, the SWAP data reveal signatures similar to the AIA data, yet with only one passband at 171~\,\AA{}, and therefore are not used in this study.

\section{Evolutionary stages of the eruption: EUV and WL observations}

According to the AIA data at 131 and 94~\,\AA{} (hot channels, note the 131~\,\AA{} channel also responds at low temperature) and (171 and 304~\,\AA{}, cool channels), the eruption can be separated into two stages (see Figure 2 and its accompanying movie). The first stage is a gradual one, initiated by the onset of the M1 flare. A highly-inclined and twisted/writhed structure (referred to as S1, see the green arrow in Figure 2(a)) is released with the flare around 11:20 UT. S1 moves slowly southeastern along an inclined path, before its eventual ejection. It is observable at 131 and 94~\,\AA{} and invisible at 171 and 304~\,\AA{}, indicating that it contains plasmas as hot as 6$\--$10 MK. It seems that one root of S1 is connected to the flaring site, and the other root is connected to a low-lying filament system (referred to as F1, see the upper blue arrow in Figure 2(c)). The F1 system gets activated along with the rising motion of S1 materials. It presents very rich dynamics, with a continuous advancing motion along the solar disk, rapid injection of brightening materials. At the same time, some structures (plasmas) are released from the F1 system into S1. This may contribute to further development and growth of S1. After 12:25 UT, S1 starts to take off at a faster pace. This starts the second stage.

During this second stage, the eruptive structure S1 becomes very diffusive and fades into the background after 12:35 UT at 131 and 94~\,\AA{}. At 171~\,\AA{}, a narrow knot$\--$like loop structure (referred to as L1, see the red arrow in Figure 2(e)) appears after 12:32 UT. Its expansion then becomes the dominant eruptive signature in the AIA FOV. From the movie of composite 131 and 171~\,\AA{} data, L1 and S1 are closely related to each other, suggesting that the structures observed in different channels correspond to the same complex ejecta. As seen from the 304~\,\AA{} images, on the southern part of F1 there exists another filament structure (referred to as F2, see the lower blue arrow in Figure 2(c)). At 304~\,\AA{}, F2 manifests as a low$\--$lying arcade of filament structures, its spine is observed as the column-like dark absorption feature at both 131 and 171~\,\AA{} (see Figures 2(a) and 2(d)).

The 304~\,\AA{} arcade of F2 starts to rise slowly around 12:35 UT. Several minutes later, the arcade becomes more and more cusp$\--$like. Around 12:45 UT, there appear bidirectional flows on the northern flank of the rising arcade (best seen from the movie accompanying Figure 2). In the meantime, F2 starts to get brightened along its outer border. The earliest brightening signature is observed at $\sim$ 12:35 UT. On its southern flank, there presents a very clear untwisting motion at 131, 171, and 304~\,\AA{} (see Figure 2 and the accompanying movie). The cusp reaches the highest altitude in the time range of 12:50$\--$12:55 UT. This sharp cusp feature and relevant dynamics indicate the presence of a current sheet atop the cusp and the occurrence of reconnection high in the corona. After that, the cusp moves sunwards and becomes dome-like.

The event is also observed by both STEREO spacecraft (SA and SB). In the SA FOV, the event is a limb one with the sources observable on the disk, while in the SB FOV the event is a backside one. The eruptive signatures observed by EUVI and COR1/2 from the twin spacecraft are very similar. Therefore, only the data from SA are presented (in Figure 3).

The structures L1, F1, and F2 are also observed by SA (see Figure 3). Again, F2 is observed as a column$\--$like dark absorption feature (at 195~\,\AA{}, pointed at with white arrows in panels b and c), embedded within a coronal cavity. From the SA perspective, the interaction process between the expanding L1 (see the red arrows in panels b and c) and the F2 system is clearly observed. Brightening features appear atop F2 around 12:35$\--$12:40 UT, and F2 is disrupted after being swept by the expanding L1 structure. Around 12:55 UT, a new EUV structure appears and points outward from inside of the cavity (see the white arrow in Figure 3(d) and the accompanying movie). The EUV structure seems to be stretched outward by the ejecta.

The above EUV data from both SDO and STEREO indicate that strong interaction takes place between the ejecta and F2. The interaction may involve reconnection. This is supported by the above$\--$presented SDO and STEREO observations, including the brightening features close to F2, bidirectional flows within the interaction region between the eruptive structure and the F2 system, untwisting motion along the southern flank of the F2 arcade, the presence of the new EUV structure that is within the F2 system and stretched outward by the eruption (indicating a topology change), and the clear cusp-like structure transformed from a smooth arcade associated with F2. The onset of the F2-associated reconnection is around or minutes before 12:35 UT, and may last for a few tens of minutes.

In the outer corona, the eruption is observed by coronagraphs (see the lower panels of Figure 3) including the LASCO C2$\--$C3, and the STEREO A/B COR1$\--$2. We see that the CME front (red arrows in panels e and f) is well above the L1 loop (the yellow arrow in panel e). After the CME propagates further away, it manifests an overall bulb-like morphology (see the LASCO C2$\--$C3 images). A bright core appears at the inner part of the ejecta. The structure between the core and the bright front is relatively dark, indicating the presence of a cavity structure. The linear CME speed (of the front) is about 770 km s$^{-1}$ according to the online CDAW catalogue, and the center of the bright core moves at a slower speed of 450 km s$^{-1}$. Both speeds are measured within the LASCO C2 FOV. The data gap between the EUV imaging region in the inner corona and the white$\--$light imaging region in the outer corona is obvious and given by the dark region in the composite image of AIA and LASCO-C2 data (Figure 3(g)). As mentioned earlier, this data gap makes it difficult to identify the correspondence between the eruptive EUV structure and components of the white-light ejecta.

\section{The T-IVm source imagery of NRH}
As seen from Figure 4 and the accompanying movie, the burst appears at 327 MHz and 298 MHz since 12:31 UT, and extends to 150 MHz since 12:35 UT. For a quantitative description of the source evolution, we show values of $T_{Bmax}$ on each NRH image, which are obtained in the domain excluding the solar disk to remove irrelevant emission from another active region (AR12087) on the disk.

At $\sim$ 12:45 UT, the source disappears at 327 MHz, and the highest t-IVm frequency shifts to 298 MHz ($\sim$ 1.43 R$_\odot$, the centroid location in heliocentric distance). Around this time, $T_{Bmax}$ increases monotonically with decreasing frequency, from 3 MK at 298 MHz to 70 MK at 173 MHz ($\sim$ 1.53 R$_\odot$) and 135 MK at 150 MHz ($\sim$ 1.56 R$_\odot$). The highest $T_{Bmax}$s are achieved in the time range from 12:50 UT to 13:00 UT, reaching up to $1\times 10^9$ K. After 13:00 UT, the source at 228 MHz vanishes, and only the sources at 173 MHz and 150 MHz persist. After 13:05 UT, only the 150 MHz source survives,which propagates up to $\sim$ 2.2 R$_\odot$. During the t-IVm burst, NRH sources present an overall spectral drift toward lower frequencies and a motion propagating away from the Sun. This is consistent with the ORFEES dynamic spectral data and confirms the presence of the t$\--$IVm burst.

According to Figure 5(a), $T_B$ tends to increase with declining frequencies. Figure 5(b) presents the spectra of $T_B$ at three moments (12:43-12:45 UT) when the t-IV bursts are observed at all relevant NRH frequencies. The three spectra present similar broken-power$\--$law like shape, being harder from 150 MHz - 228 MHz and softer at higher frequencies. The spectral indices are around $-3$ for the pre-break spectra and around $-9$ for the spectra above the break. Panel c of Figure 5 presents the polarization measurement, which is obtained by taking average over the area within the 85{\%} contour of $T_{Bmax}$. We see that the polarization levels of the t$\--$IVm burst remain in-general weak ($<\sim$ 10\%) at most observational times and frequencies.

The source speed can be estimated by measuring the motion of the source at a fixed frequency. From 12:35 to 13:05 UT, the speed of the t$\--$IVm source is 185 km s$^{-1}$ at 270 MHz, 253 km s$^{-1}$ at 228 MHz, 330 km s$^{-1}$ at 173 MHz, and 355 km s$^{-1}$ at 150 MHz. Note that the frequency of t-IVm sources changes with their outward motion. This should be taken into account when estimating the source speed. To do this, in Figure 1, we delineated the upper bound and the lower bound of the t$\--$IV spectrum. By averaging them, we get the central band of the burst. Picking out those NRH sources that correspond to the time-frequency values of this central band, the overall t-IVm burst source speed can be inferred. This is done by linearly fitting the source centroid positions along the central band at relevant NRH frequencies obtained at subsequent times. The fitted speed is about 360 km s$^{-1}$, slightly larger than the value obtained for those at fixed frequencies. This speed will be further used to extrapolate the source location to lower frequencies (and later times).

\section{Combined analysis of different data sets}

In Figure 6(a-d) we plot composite images of radio sources and 94~\,\AA{} data of AIA. The temporal variation of the composite data has been presented in the accompanying movie. It can be seen that the radio sources are located at the top front of the bright core of the ejecta that is the high-temperature structure observed at 94 and 131~\,\AA{} (i.e., the S1 structure). Another important feature of radio sources is their spatial dispersion, i.e., sources at different frequencies do not overlap with each other. Instead, these sources are dispersed spatially, together they manifest a nice formation with an arcade-like morphology (see Figure 6(c-d)), with sources at higher frequencies being closer to the Sun.

After 12:50 UT, most t$\--$IVm sources at lower NRH frequencies propagate beyond the AIA FOV. The last observable t$\--$IVm source (at 150 MHz) presents at $\sim$2.2 R$_\odot$ and 13:05 UT. This already extends into the LASCO C2 FOV.
Thus, it is clear that the NRH data at different frequencies can be combined to track the source (and the corresponding CME component) continuously from the inner to the outer corona and thus fill in the gap between AIA and LASCO-C2 data.

To further reveal the counterpart of t-IVm sources in the WL CME, in Figure 7(a) we show one composite image of the last 150~MHz source and the C2 data at 13:12 UT. The CME presents a double-front morphology, with the external front likely given by the material piled-up by the ejection and the internal front given by the ejecta itself. Note that the C2 image is about 7 minutes later than the last radio source at 150 MHz. This means we have to extrapolate the location of the radio source from 13:05 UT to 13:12 UT. Assuming a constant speed of the source at 360 km~s$^{-1}$, within 7 minutes it can propagate outward by $\sim$ 200". As seen from Figure 7(a), the source, if being extrapolated to 13:12 UT, shall be located at the upper front of the bright ejecta.

Using the central band of the t$\--$IVm burst given in Figure 1 (see the dashed line on the dynamic spectrum), we can estimate the central frequency of the t-IV burst to be around 40 MHz at 13:17 UT and around 20 MHz at 13:48 UT. We further extrapolate the location of the radio sources from that observed at 13:05 UT and at 150 MHz to 40 MHz (13:17) and at 20 MHz (13:48). The results are shown in Figures 7(b) and 7(c). Again we see that both extrapolated sources are located at the upper front of the internal bright ejecta, consistent with the conclusion deduced for 150~MHz. This reveals the counterpart of the eruptive structure observed at 131~\,\AA{} in the inner corona to that observed in the outer corona with LASCO C2.

\section{DISCUSSION}
We first summarize major similarities and discrepancies between the present event (20140615) and that reported earlier (20120304) by Vasanth et al. (2016). Both events present a relatively high $T_B$ in the range of 10$^7$ - 10$^9$ K and significant source dispersion with observing frequencies, and both are associated with hot plasmoid-like structure observed by AIA. Major differences lie in the following aspects. (1) The present event occurs during the declining stage while the earlier one occurs around the impulsive stage of the flare. The above two differences are significant when considering the source of energetic electrons. (2) The present event lasting for more than two hours, much longer than the earlier event which lasts for about 20 minutes. (3) The polarization level of the present event is in-general weak while the earlier event manifests a moderate level with a change of polarization sense from negative to positive. And, (4) the source at 150 MHz in the present event propagates outward to a distance of $\sim$2 R$_\odot$ while in the earlier event the source at 150 MHz fades away around the outer border of the AIA FOV. The large source height of the present event allows us to trace the T-IVm sources from the AIA FOV all the way to the LASCO C2 FOV.

The analysis of the last section reveals the potential of using the t$\--$IVm source imagery to bridge the gap of EUV data in the inner corona and the WL data in the outer corona. It is found that the t-IVm sources correspond to the upper part of the EUV hot eruptive structure in the inner corona and to the same part of the bright WL CME ejecta in the outer corona. This indicates that the bright CME core is originated from the hot EUV structure. Most earlier studies suggest that the bright core corresponds to eruptive filament, while latest studies suggest that this may not be the case for some events (Howard et al. 2017; Song et al. 2017). In our studies, no signatures of eruptive filament are observed although the event is associated with strong filament activities. Therefore, our study supports that at least a part (if not all) of the CME bright core is associated with the hot EUV eruptive structure.

In Section 3, we have presented several pieces of evidence of magnetic reconnection induced by the interaction between the eruptive structure and nearby filament system. As seen from the GOES SXR light curves, during the onset ($\sim$ 12:35 UT) and the first few tens of minutes of this reconnection process, the SXR profile is in the declining phase and reaches a local minimum. This is likely due to the relatively high reconnection site, in comparison to the usual flaring reconnection, thus not causing any observable X$\--$ray emission. In addition, it seems unlikely for the M1$\--$class flare peaking at 11:45 UT ($\sim$ 50 minutes earlier than the t-IVm burst onset) to directly supply the t-IVm emitting electrons. The onset time of the post-flare high-corona reconnection and the start time of the t$\--$IVm burst are well consistent with each other. Therefore, the observations support that the radio-emitting electrons are likely supplied by this reconnection process. From the AIA data, the hot eruptive structure is highly$\--$twisted, this may create sites of local magnetic enhancement and favor the confinement of energetic electrons through the well$\--$known magnetic mirror effect (see also Wu et al. 2016 and Vasanth et al. 2016).

The t-IVm burst (and the associated energetic electrons) lasts for at least 2.5 hours. This spans the essential stage of CME dynamics from the inner to the outer corona. It should be highlighted that the above-mentioned bridging role of t-IVm source imaging data is significant since this provides a means to visualize one essential component of the ejecta. According to our study, this component corresponds to the high-temperature twisted structure, likely a flux rope that acts as energy carrier and driver of the eruption. In addition, it is the container of those radio-emitting energetic electrons, thus it may be of special interest to studies on how energetic electrons are transported from near the Sun to the interplanetary space. The long duration of the t$\--$IVm burst indicates that the Coulomb collision of energetic electrons with background plasmas is not important, and their escape from the magnetic trap, if any, does not extinguish the radio emission. Yet, it is not known whether the energetic electrons can survive till the CME arrival near 1 AU.

The emission mechanism of the t$\--$IVm burst still remains controversial (see the introduction section). Our data show that the t$\--$IVm burst of study has a clear spatial dispersion over different frequencies, and $T_B$ can be as high as $\sim 10^9$ K. These observations strongly support that the underlying emission mechanism is coherent. Other properties that are useful to constrain the emission mechanism include the in$\--$general weak polarization level (at all relevant frequencies), the broken power$\--$law spectra of the emission intensity. The weak polarization level can be explained by the presence of twisted magnetic structure in the source region. The very-large source height reported here puts a challenge to the plasma emission mechanism since this corresponds to a density of $\sim$3$\times$ 10$^{8}$ cm$^{-3}$ for fundamental plasma emission (at 150 MHz) and 7$\times$ 10$^{7}$ cm$^{-3}$ for harmonic around 2.2 R$_\odot$. The obtained density values are much higher than those expected from coronal density models such as the Newkirk or the Saito model (Newkirk 1961; Saito 1970). Fortunately, the multi-wavelength EUV data of AIA provides a way to infer the density values of the t-IV source and thus to further examine the plasma hypothesis.

First, it is not surprising to have a much higher density than the quiet background of the corona since the t-IV source is associated with a dense eruptive structure. Keeping this in mind, we conduct differential emission measure (DEM) analysis of the eruptive structure at about 6-7 minutes before the t-IV presence so as to have enough photons for the analysis. We select two moments (12:24 and 12:25 UT) using the method developed by (Schmelz et al. 2011; Cheng et al. 2012). The results are shown in Figure 8. Since the eruptive structure presents a gradual expansion from 12:24 UT to 12:31 UT (see the movie accompanying Figure 2), we expect that the density of the eruptive structure at 12:31 UT is smaller yet still reasonably close to the density around 7 minutes ago. The EM values of the EUV structure at 12:24 UT and 12:25 UT are found to be around 1$-$2$\times$10$^{27}$~cm$^{-3}$, and the temperature is $\sim$5$-$8 MK. Assuming the depth of the structure is approximately equal to its width ($\sim$120"), we infer that the density of the EUV structure is around 3$-$6$\times$10$^{8}$~cm$^{-3}$ at 12:24-12:25 UT. This corresponds to a frequency in the range of 160$-$220 MHz for fundamental plasma radiation and 320$-$440 MHz for harmonic plasma radiation. The latter is closer to the highest frequency (327 MHz according to NRH) at the start of the t-IV burst ($\sim$12:31 UT). Yet, due to the uncertainty of the DEM method, and the two facts that (1) the data used for DEM are obtained $\sim$7 minutes before the t-IV appearance, and (2) the t-IV sources are co-spatial with the diffuse top part of the EUV structure where the density might be smaller than the central part, the above statement should be taken with caution.

The observations of significant source dispersion with observing frequency and the relatively high $T_B$ ($<\sim$10$^9$~K) do not favor gyro-synchrotron to be the radiation mechanism. In addition, with the large density as inferred from the DEM analysis, the Razin suppression effect becomes serious for gyro-synchrotron radiation. For example, assuming that the density is $\sim$10$^8$~cm$^{-3}$ and the magnetic field strength is about 5 Gauss at 2 R$_\odot$, the Razin frequency is then $\sim$380 MHz and will be higher if considering a more-likely weaker field. This is inconsistent with the observed spectra that indicate the Razin frequency is less than 150 MHz (see Figure 5b). Using Ramaty's equations (Ramaty 1969) and the corresponding code distributed with the SolarSoftWare (SSW) package, we did simulate the gyrosynchrotron emission assuming a reasonable power-law spectrum of energetic electrons and the above values of density and magnetic field strength. The obtained emission is much weaker than observed and the observed spectra cannot be reproduced.

The imaging observations at different wavelengths suggest that the radio burst is excited by energetic electrons that are trapped within the eruptive structure. Thus, we suggest that the loss$\--$cone instability driven by electron population inversion with respect to the perpendicular component of velocity (i.e., ${\partial f \over \partial v_{\perp}} > 0$) is relevant. The instability may excite electron cyclotron maser emission (ECME) if the electron cyclotron frequency $\omega_{ce}$ is much larger than the plasma oscillation frequency $\omega_{pe}$ (Wu {\&} Lee 1979; Lee {\&} Wu 1980). However, in the outer corona beyond 1.5 R$_\odot$ it is expected that $\omega_{pe}$ is larger than $\omega_{ce}$. Thus, ECME may not be the mechanism for the t-IVm burst of study. In this parameter regime, Winglee et al. (1986) showed that the loss-cone instability may drive the z mode which further converts to escaping radio emission. Winglee et al. (1986) also demonstrated that the z-mode emission can account for the wide-band continuum like the t-IV solar radio burst. These studies used simplified electron velocity distribution functions with homogeneous magnetic field and plasma parameters in the background. Further studies shall consider the observational constrains provided here, in particular, on the source properties, to achieve a better understanding of the emission mechanism.

\section{SUMMARY}
Along the Sun-Earth line, there exists an imaging data gap between CME observations at EUV in the inner corona and those at WL in the outer corona (with space-based instruments). This makes it difficult to identify the correspondence of magnetic structures in different regimes and to investigate their dynamical and morphological evolution. Here we present a case study on a t$\--$IVm burst with combined analysis of different data sets. One major purpose is to highlight the potential of using t$\--$IVm source imageries to track continuously the eruptive structure, in particular the specific component containing radio-emitting energetic electrons, and thus somehow bridge the data gap. Another purpose of our study is to infer the source parameters as well as the underlying emission mechanism by taking advantage of the state-of-the-art multi-wavelength AIA/SDO data.

The sources of the t$\--$IVm burst are associated with a hot highly$\--$twisted structure observed at 131 and 94~\,\AA{} by AIA. The structure is connected with the interaction region between the eruptive structure and a nearby filament system. Within this region, significant evidence of magnetic reconnection is observed. The reconnection may have generated energetic electrons that are injected into and trapped by the eruptive structure. These electrons then excite the wide-band moving type-IV burst. The sources first appear in the AIA FOV, corresponding to the upper part of the hot AIA eruptive structure. They propagate outward consistently, while the EUV structure fades out in the middle part of AIA FOV. Later, t$\--$IVm sources propagate into the LASCO C2 FOV, also correspond to the upper part of the bright core of the CME ejecta. This suggests the close association of the 131~\,\AA{} eruptive structure and the bright core of the CME.

We find that the t$\--$IVm burst sources present a clear spatial dispersion with frequencies, and they line up together to form a section of an arcade$\--$like structure. The burst lasts for at least 2.5 hours from 327 MHz to the lowest observable frequency of the NDA spectrometer ($\sim$ 10 MHz) or even lower frequency (no data available). In addition, the $T_B$ t-IVm burst varies from 10$^7$ to 10$^9$ K, increasing generally with declining frequency. This leads to a broken power$\--$law type of spectra. The polarization levels of the burst remain weak in general. As summarized above, energetic electrons accounting for the burst are likely accelerated via reconnection that takes place relatively high in the corona, and then injected into and trapped by the eruptive structure. They remain radio$\--$loud in the whole 2.5$\--$hour long duration of the t$\--$IVm burst.

DEM analysis of AIA data shows that the electron density within the corresponding source structure is at the order of $10^8$ cm$^{-3}$. Calculations suggest that the gyrosynchrotron emission may not be of interest here, and our studies support that the T-IV burst is given by plasma emission associated with trapped electrons. One likely process to excite the plasma emission is through the z-mode maser instability. According to the density measurement with AIA, the instability shall grow in the parameter regime of $\omega_{pe} >> \omega_{ce}$ and the temperature of the background plasma may be as high as several MK. The study thus provides further constraints on source conditions of T-IVms. Further studies on the emission mechanism are demanded to explore the full diagnostic potential of the radio data presented here.

\acknowledgements We thank the ORFEES, DAM, and NRH teams for providing the high-quality radio spectral and imaging data. SDO is a mission of NASA¡¯s Living With a Star Program. The STEREO/SECCHI data are produced by an international consortium of the NRL, LMSAL and NASA GSFC (USA), RAL and University of Birmingham (UK), MPS (Germany), CSL (Belgium), IOTA, and IAS (France). SOHO is an international collaboration between NASA and ESA. This work was supported by NNSFC grants 41331068, 11790303 (11790300), and NSBRSF 2012CB825601.

\begin{figure}
\includegraphics[trim = 22mm 170mm 10mm 2mm, clip]{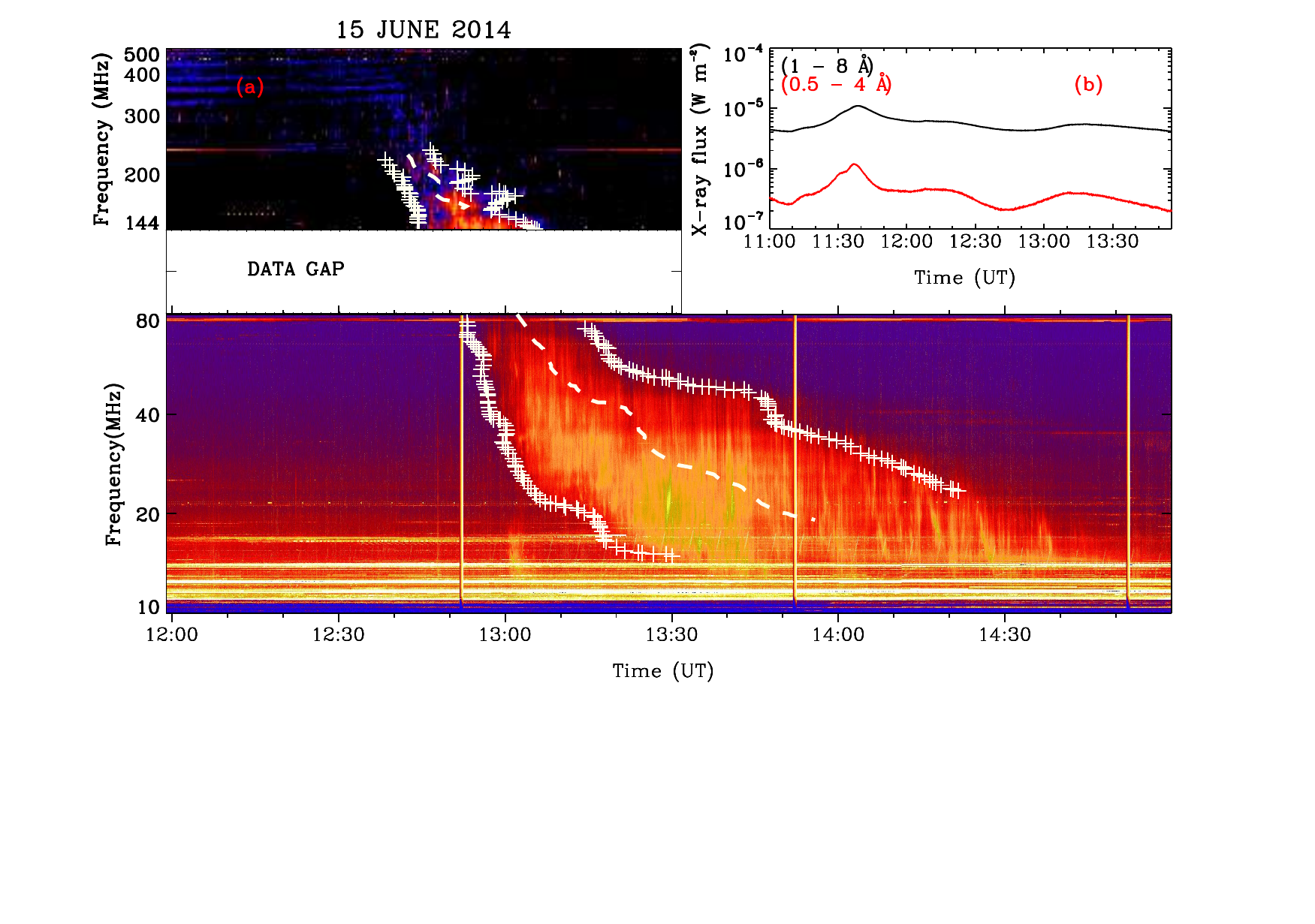}
\caption{The composite of the dynamic spectra of the type-IV burst recorded by ORFEES (300$\--$144 MHz, upper part) and NDA (80$\--$10 MHz, lower part). A data gap exist between 144 MHz and 80 MHz. The plus signs show the upper and lower borders of the dynamic spectra. The average of the values on the borders is given by the dashed line plotted in panel a, representative of the central band of the burst. (b) The GOES SXR light curves (0.5-4 \& 1-8~\,\AA{}). }\label{Figure 1}
\end{figure}

\begin{figure}
\includegraphics[trim = 15mm 100mm 25mm 20mm, clip]{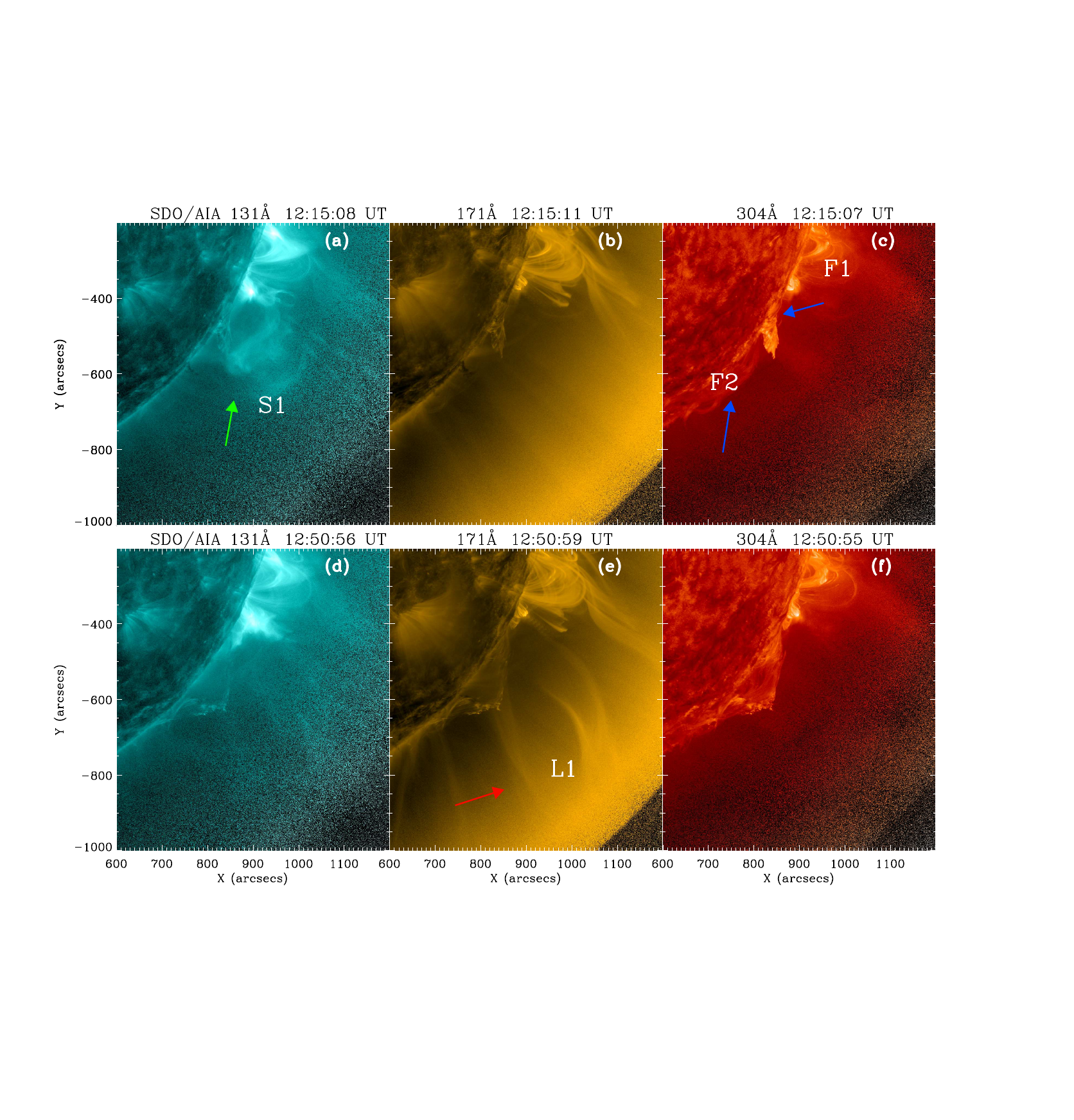}
\caption{AIA images of the event at 131 (a and d), 171 (b and e), and 304 (c and f)~\,\AA{} at two times (12:15 UT and 12:50 UT), to show the eruptive structures (S1 and L1) and relevant filament (F1 and F2) activities. An animation of the AIA images is available. The animation runs from 11:20 to 13:04 UT.}\label{Figure 2}
\end{figure}

\begin{figure}
\includegraphics[trim = 20mm 100mm 15mm 20mm, clip]{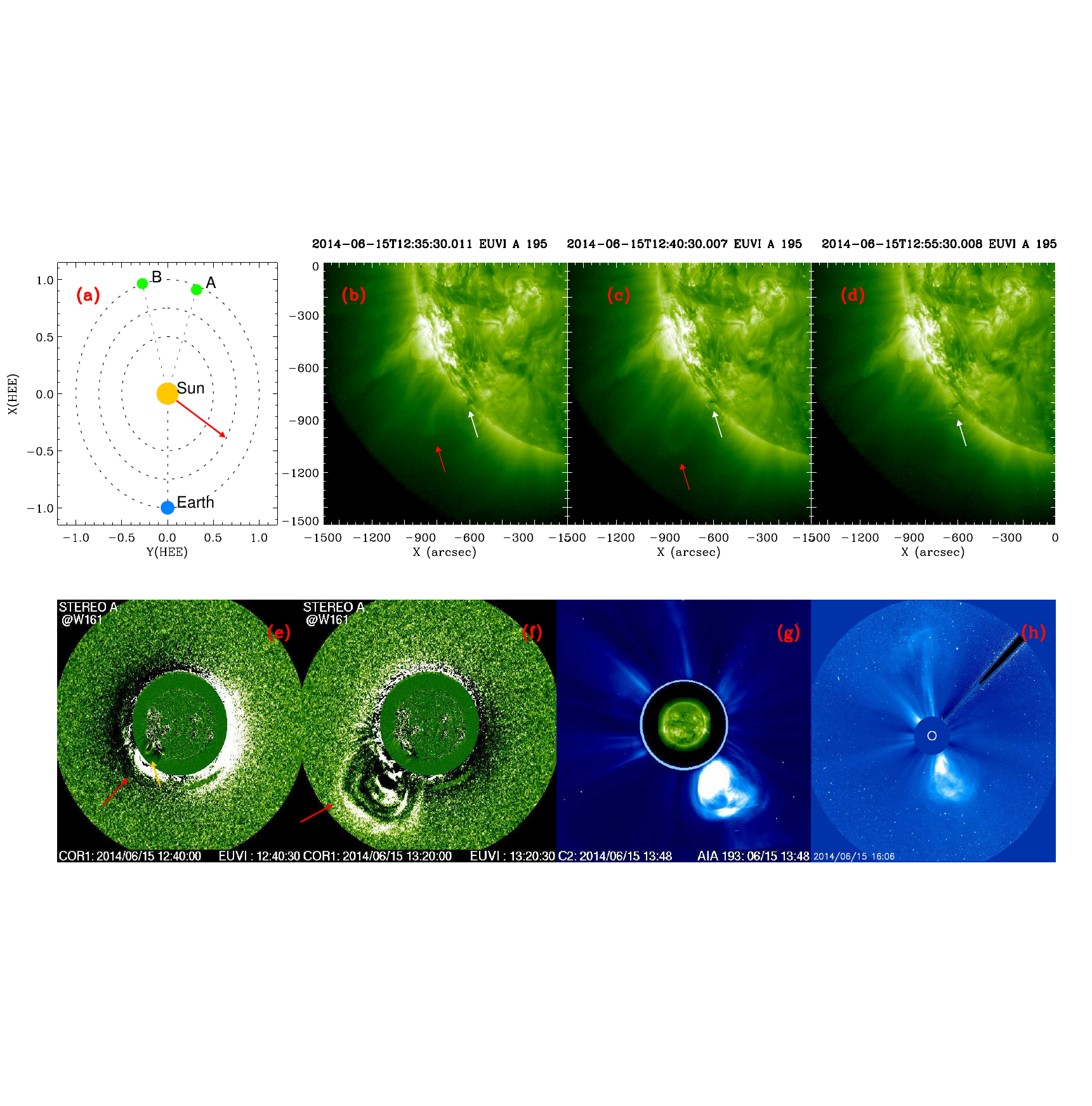}
\caption{(a) The relative location of the twin \emph{STEREO} spacecraft and the Earth (\emph{SDO}). (b-d) Sequence of EUVI/SA images showing the eruptive structure at 195~\,\AA{}. (e-h) The CME structure observed by COR1/SA at 12:40 and 13:20 UT, and by LASCO-C2 at 13:48 UT and by C3 at 16:06 UT. The CME presents an overall bulb-like morphology with a bright core. An animation of the EUVI/SA  195~\,\AA{} images is available. The animation runs from 10:45 to 14:00 UT in 5 minute steps and includes the same annotations as the static version.} \label{Figure 3}
\end{figure}

\begin{figure}
\includegraphics[trim = 2mm 5mm 15mm 170mm, clip]{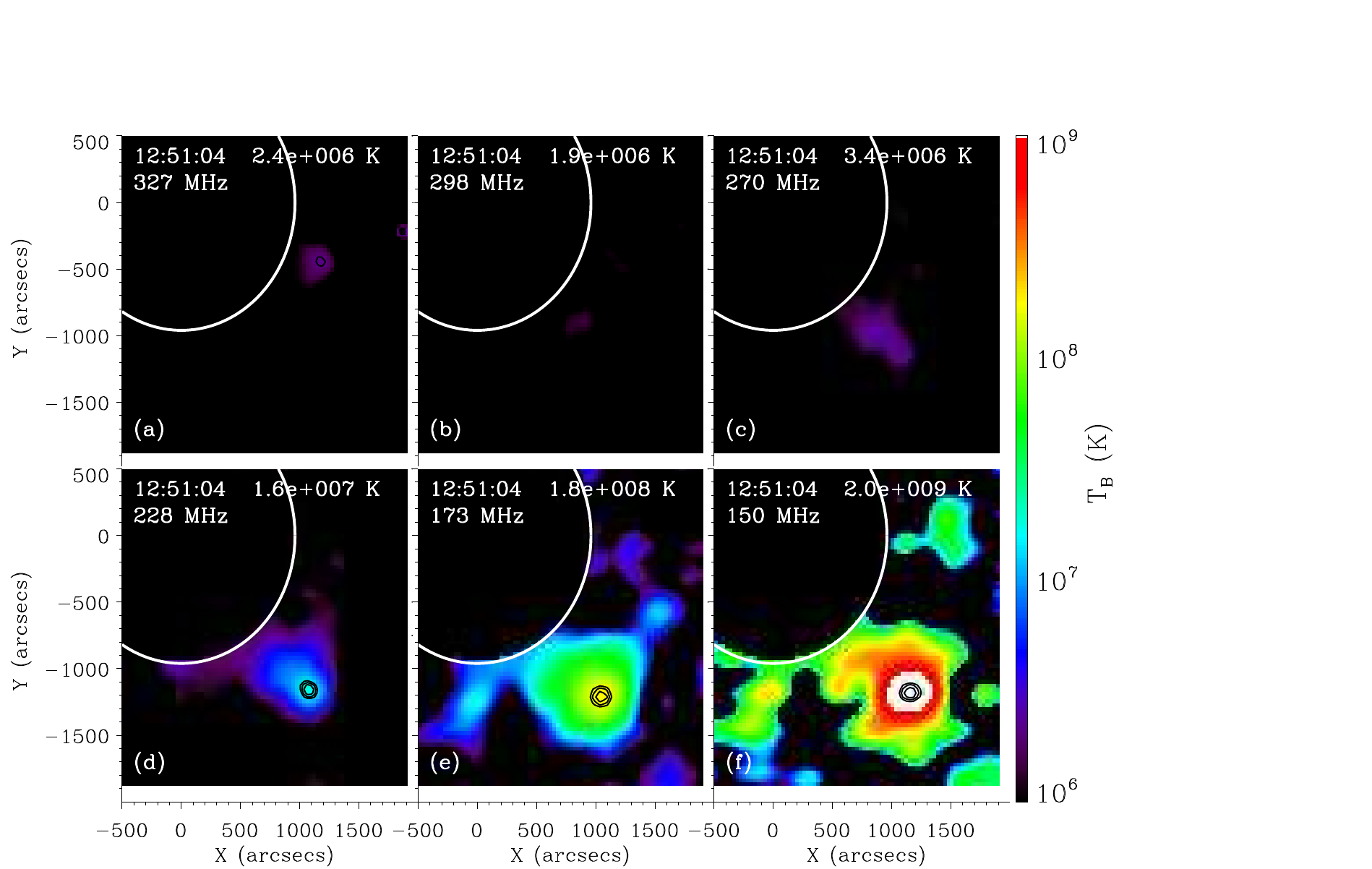}
\caption{The type-IV sources observed by NRH at different frequencies at 12:51 UT. The maxima of $T_B$ are shown in each panel. The contours are representative of 85, 90 and 95 \% of corresponding $T_{Bmax}$. An animation of this figure is available. The animation runs from 12:15 to 13:15 UT. }\label{Figure 5}
\end{figure}

\begin{figure}
\includegraphics[trim = 0mm 0mm 15mm 245mm, clip]{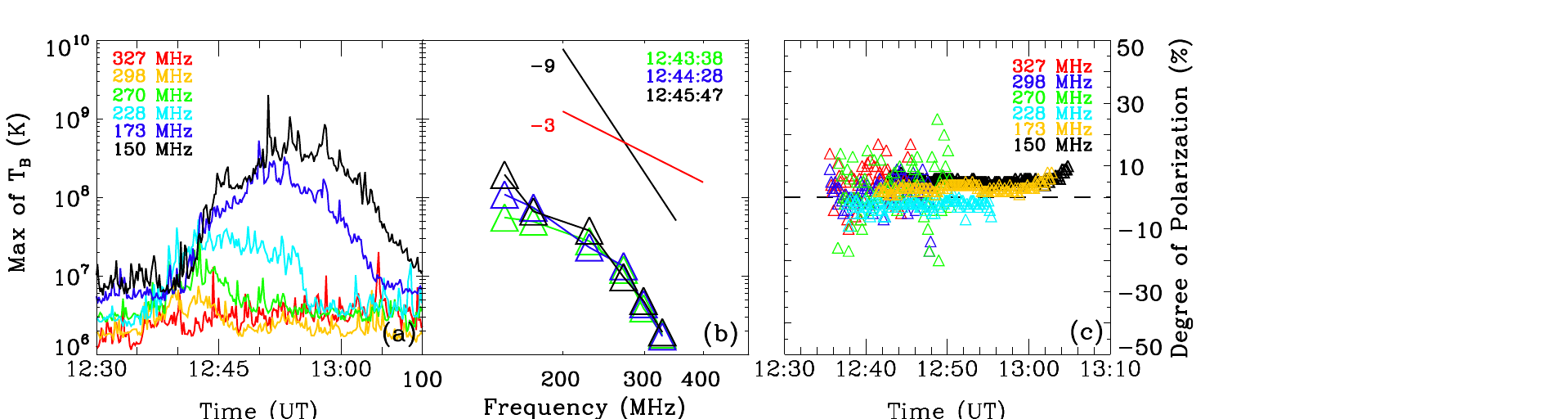}
\caption{Temporal variations of $T_B$ at different frequencies (a), the $T_B$ spectra measured at three selected moments (b), and polarization levels at relevant frequencies (c).}\label{Figure 5}
\end{figure}

\begin{figure}
\includegraphics[trim = 2mm 0mm 15mm 110mm, clip]{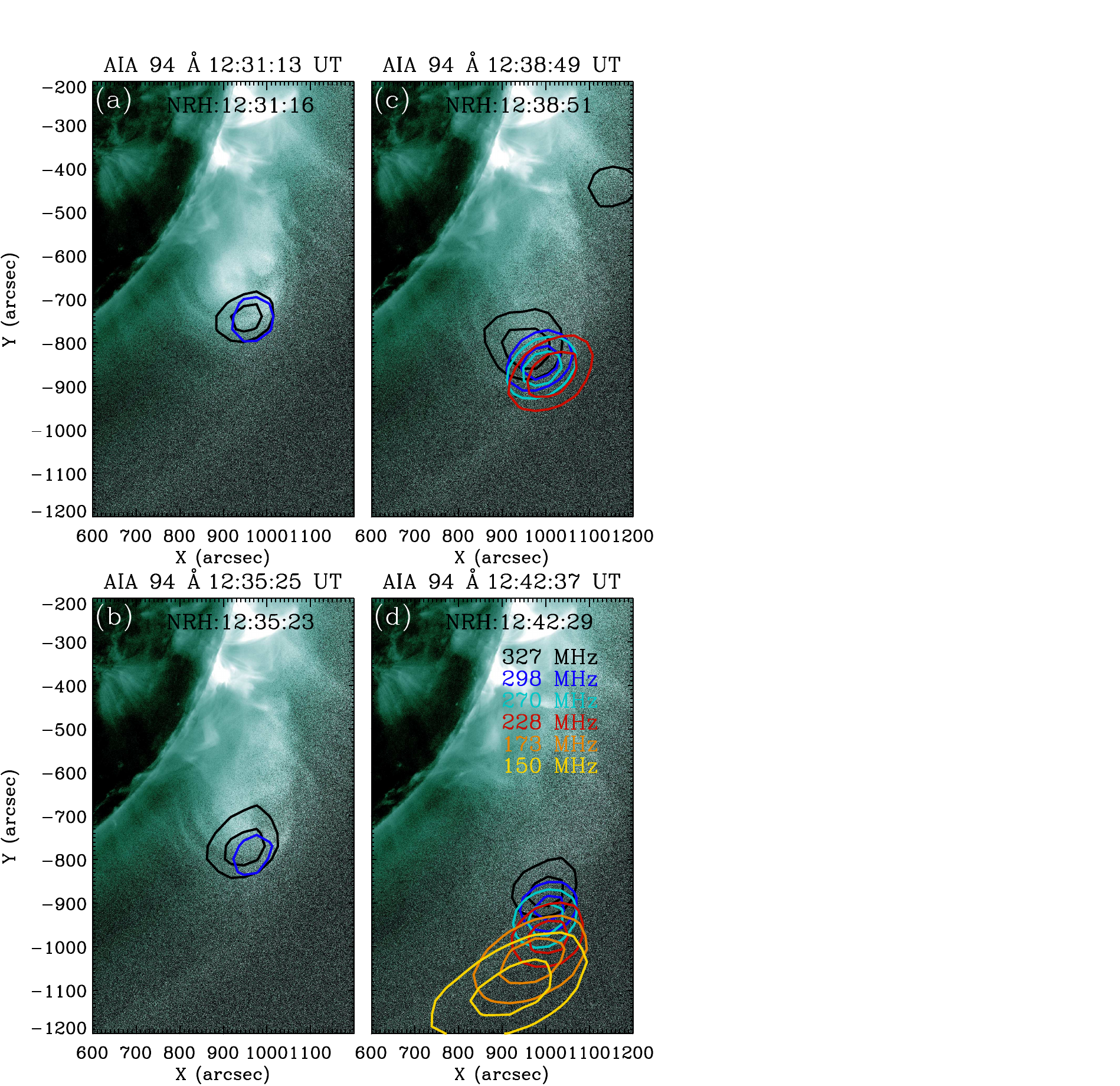}
\caption{Type-IV sources observed by NRH at four moments superposed onto the AIA-94~\,\AA{} images observed at closet times. NRH sources are represented by 85 and 95 \% contours of corresponding $T_{Bmax}$. An animation of this figure is available. The animation of the Type-IV sources runs from 12:10 to 13:20 UT and covers a spatial scale significantly larger than the static version(s), tracing sources propagating beyond the AIA FOV. }\label{Figure 6}
\end{figure}

\begin{figure}
\includegraphics[trim = 20mm 100mm 0mm 50mm, clip]{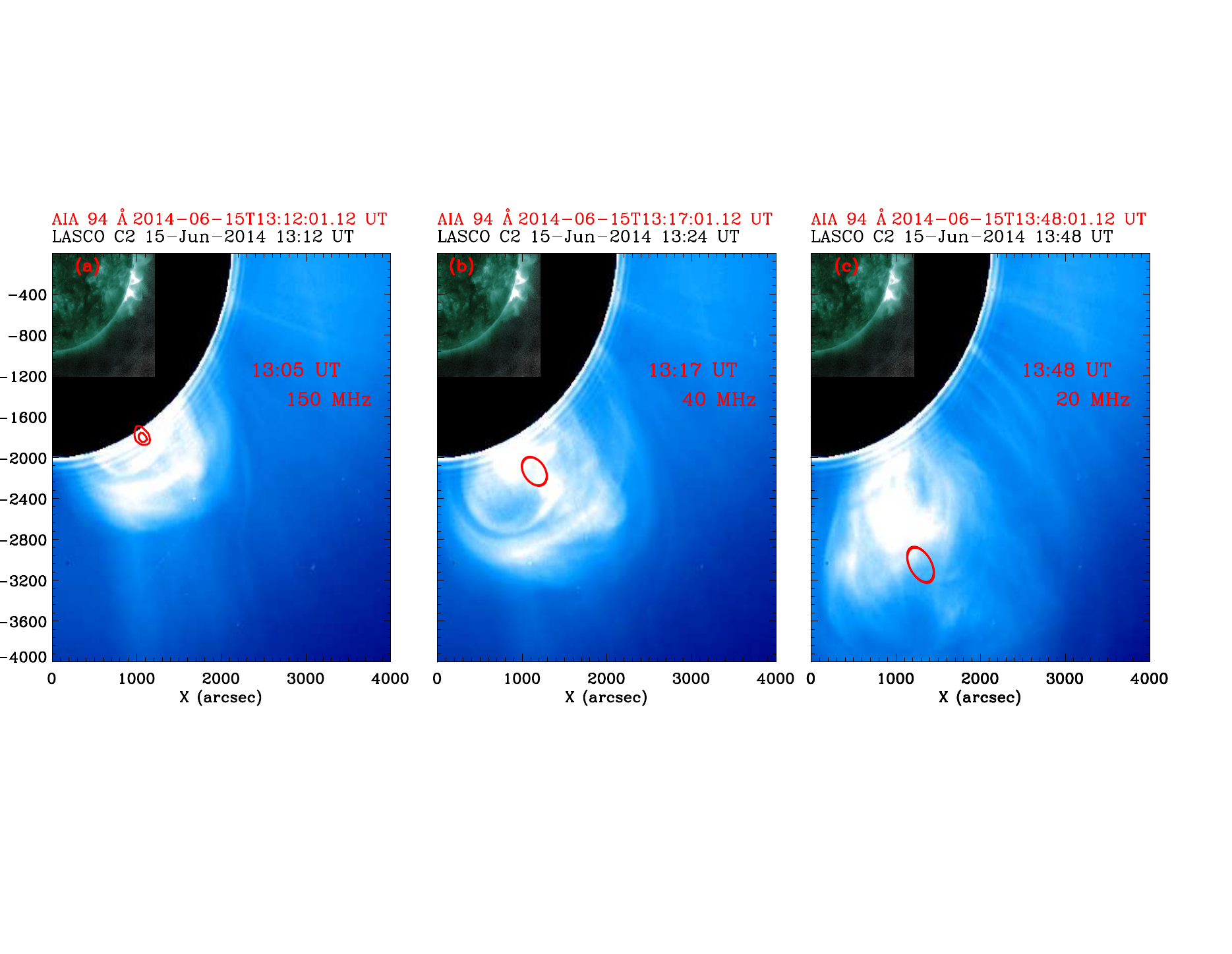}
\caption{The t-IVm source observed at 150 MHz and 13:05 UT superposed onto the LASCO-C2 images observed at 13:12 UT (a), t-IVm sources are represented by 85 and 95 \% contours of corresponding $T_{Bmax}$. The extrapolated t-IVm sources (40 MHz at 13:17 UT (b), and 20 MHz at 13:48 UT (c)) superposed onto the C2 images (see text for details). The size of the extrapolated sources is given arbitrarily.}\label{Figure 7}
\end{figure}

\begin{figure}
\includegraphics[trim = 2mm 0mm 15mm 120mm, clip]{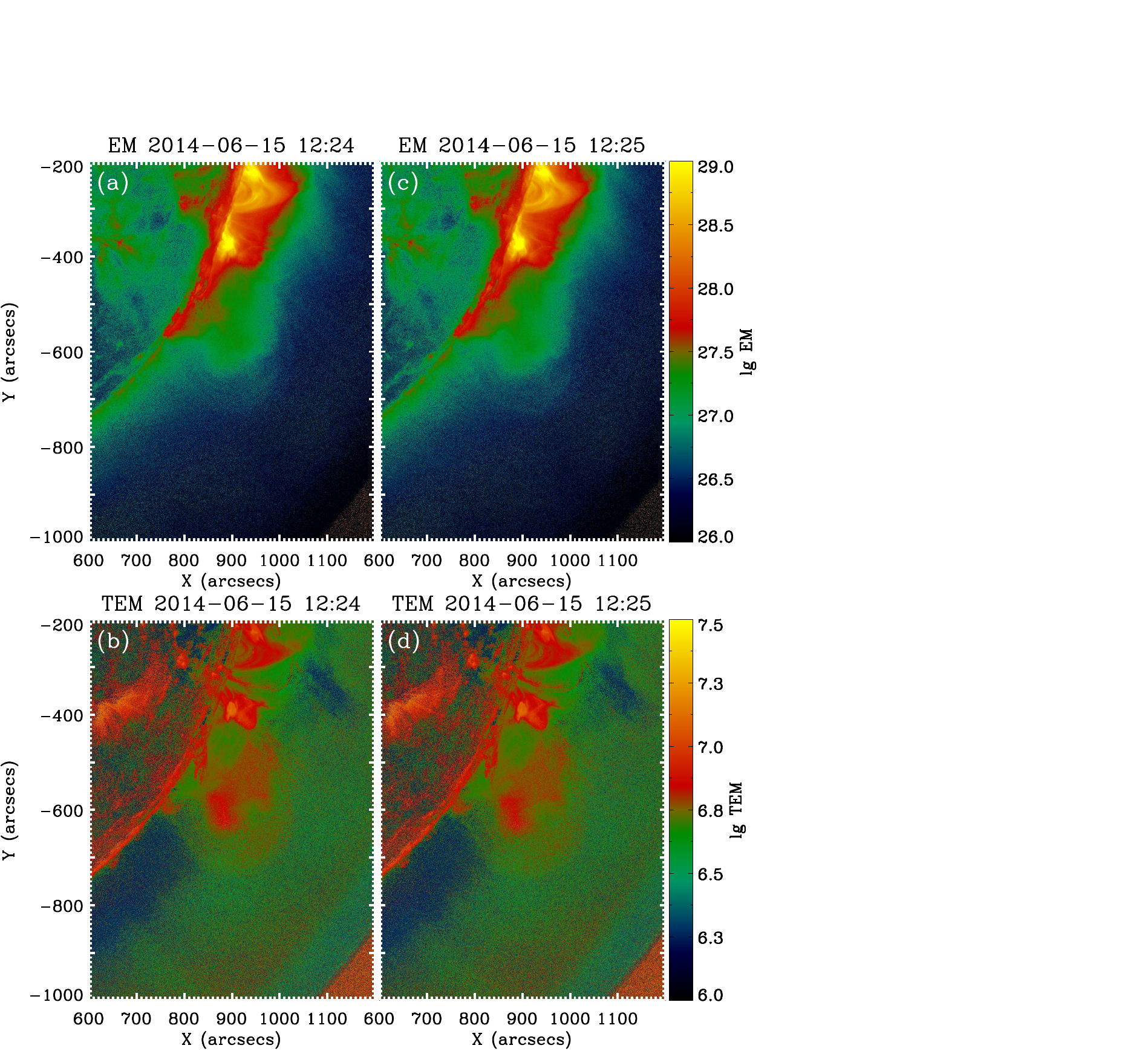}
\caption{DEM analysis results of the multi-wavelength AIA data at 13:24 UT and 13:25 UT. Upper panels are for the emission measure and lower panels are for the deduced temperature.}\label{Figure 8}
\end{figure}
\end{document}